\documentclass[12pt]{article}
\usepackage{graphicx}

\newcommand\pubnumber{Author's note number}
\newcommand\pubdate{\today}
\usepackage{graphicx}
\usepackage{hyperref}
\usepackage{epstopdf}
\usepackage{slashed}
\usepackage{rotating}
\usepackage{float}
\usepackage{array}
\newcommand{\be}{\begin{equation}}
	\newcommand{\ee}{\end{equation}}
\newcommand{\bea}{\begin{eqnarray}}
	\newcommand{\eea}{\end{eqnarray}}
\newcommand{\ba}{\begin{array}}
	\newcommand{\ea}{\end{array}}

\def\Title#1{\begin{center} {\Large #1 } \end{center}}
\def\Author#1{\begin{center}{ \sc #1} \end{center}}
\def\Address#1{\begin{center}{ \it #1} \end{center}}

\newcommand\pubblock{\rightline{\begin{tabular}{l} \pubnumber\\
         \pubdate  \end{tabular}}}
\newenvironment{Abstract}{\begin{quotation}  }{\end{quotation}}
\newenvironment{Presented}{\begin{quotation} \begin{center} 
             PRESENTED AT\end{center}\bigskip 
      \begin{center}\begin{large}}{\end{large}\end{center} \end{quotation}}

\begin{document}
\begin{titlepage}
 \pubblock
\vfill
\Title{Axial-vector form factors of the light, singly and doubly charmed baryons in the chiral quark constituent model}
\vfill
\Author{ Harleen Dahiya\textsuperscript{1},  Suneel Dutt\textsuperscript{1}, Arvind Kumar\textsuperscript{1}, Monika Randhawa\textsuperscript{2}}
\Address{\bf $^1$ Department of Physics, Dr. B.R. Ambedkar National
Institute of Technology, Jalandhar, 144027, India
\\
\bf $^2$ University Institute of Engineering and Technology,
Panjab University, Chandigarh, 160014, India}
\vfill
\begin{Abstract}
The axial-vector form factors of the light, singly and doubly charmed baryons are investigated in the framework of $SU(4)$ chiral constituent quark model. The axial-vector form factors having physical significance correspond to the generators of the $SU(4)$ group with flavor singlet $\lambda^{0}$, flavor isovector $\lambda^{3}$, flavor hypercharge $\lambda^{8}$ and flavor charmed $\lambda^{15}$ combinations of axial-vector current at zero momentum transfer. In order to further understand the $Q^2$ dependence of these charges, we have used the conventionally established dipole form of parametrization.
\end{Abstract}
\vfill
\begin{Presented}
DIS2023: XXX International Workshop on Deep-Inelastic Scattering and
Related Subjects, \\
Michigan State University, USA, 27-31 March 2023 \\
     \includegraphics[width=9cm]{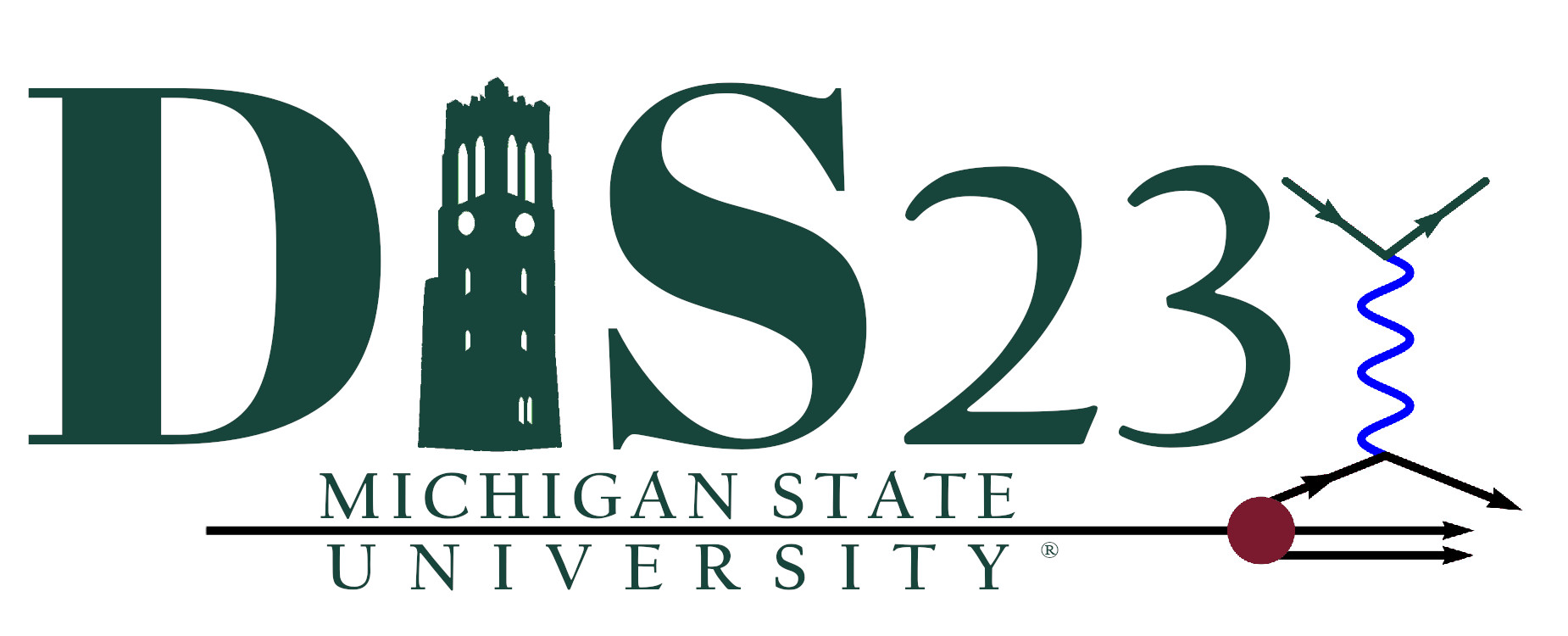}
\end{Presented}
\vfill
\end{titlepage}
\section{Introduction}

After the first direct indication for the point like constituents in the nucleon from the measurements of polarized structure functions in the deep inelastic scattering (DIS) experiments \cite{emc}, showing surprisingly that only 30\% of the proton spin is carried by the constituent quarks, a lot of experiments have been conducted to probe the structure of proton which is regarded as one of the primary goals of the upcoming Electron-Ion Collider (EIC) project.
In the context of axial-vector charges, our information on the heavy baryons is rather limited because of their short lifetimes and  it is difficult to measure their properties experimentally. Very recently, results have been reported for light and singly  heavy baryons in the chiral quark soliton model \cite{cqsm-Kim-2022}. Chiral constituent quark model ($\chi$CQM) \cite{manohar} provides  unique and important information on the distribution structure of the quarks in the baryons and is rested upon the elementary idea of chiral symmetry breaking which takes place at a much smaller distance scale as compared to that of the confinement scale. This symmetry breaking directs the massless quarks, coupled with internal Goldstone bosons (GBs), to attain dynamical mass  \cite{cheng,johan,song,hd}. This in result provides crucial indications to interpret  the nonperturbative aspects of QCD physically \cite{hds,hdcharm,nscharm,chargeradii-charm}. The purpose of the present communication is to evaluate the axial-vector charges of the light and charmed baryons in the framework of $SU(4)$ $\chi$CQM for the case of  spin $\frac{1}{2}^+$ mixed symmetry 20-plet with singly charmed sextet (6,1) as well as anti-triplet ($\bar{3}$,2) and triplet (3,2) with doubly charmed baryons.

\section{Axial-Vector Charges for the spin $\frac{1}{2}^+$  charmed baryon multiplets}

The axial-vector operator constituting the quark field for the spin $\frac{1}{2}^+$   charmed baryon multiplets can be defined  as $
A^{\mu,a}={\bf \overline{q} (x)}\gamma^\mu \gamma_5\frac{\lambda^a}{2} {\bf q(x)},$
where ${\bf q(x)}$ designates the flavor space quark field for the light and charm quarks ${\bf q}=(u,d,s,c)$. Here $\lambda^a$ ($a=1,2,..15$) are the well known Gell-Mann matrices describing the flavor $SU(4)$ structure of the light and charm quarks. A unit matrix $\lambda^0(=\sqrt{\frac{2}{3}}I)$ can be conveniently introduced in addition to these
matrices giving the axial-vector charge corresponding to the flavor singlet current $g^{0}$. In the present context of axial-vector charges, we can construct only those matrices from the generators of $SU(4)$ gauge group which have diagonal representation. We have the generator $\lambda^3$  corresponding to the flavor isovector (triplet) current,  $\lambda^8$ corresponding to the flavor hypercharge axial-vector (octet) current and the other diagonal generator $\lambda^{15}$ which respectively give the axial-vector charges as $g^{3}$, $g^{8}$ and $g^{15} $ \cite{pcac-ref1,pcac-ref2}. 
 The axial-vector current
\be \langle
B^{\frac{1}{2}^+}(p', J_z')|A^{\mu,a}|B^{\frac{1}{2}^+}(p,J_z)\rangle=\bar u(p', J_z') \left[
\gamma^\mu \gamma_5 G^a_A(Q^2)+ \frac{q^\mu}{2M_B} \gamma_5 G^a_P(Q^2) \right]  u(p, J_z)\,, \label{Amu12} \\
\ee where  $M_B$ is the baryon mass. The Dirac spinors of the initial (final) baryon states are presented as  $u(p)$ ($\bar u(p')$) and the four momenta transfer as $Q^2 = -q^2$, where $ q \equiv
p- p'$.  The functions $G^a_A(Q^2)$ and $G^a_P(Q^2)$ ($a=0,3,8,15$)
are the axial-vector and the induced pseudoscalar form factors respectively.

\section{SU(4) chiral constituent quark model and spin polarizations for spin$-{\frac{1}{2}}^+$  charmed baryon multiplets}
\label{cccm} In the $\chi$CQM, the basic process is the
internal emission of a Goldstone Boson (GB) by a constituent quark in the light and charmed baryons. This GB which consists of a 15-plet and a singlet further splits into a $q \bar q$ pair as $ q_{\pm}
\rightarrow {\rm GB}^{0}+q'_{\mp} \rightarrow (q \bar
q^{'})+q'_{\mp}\,.  $
This $q \bar q^{'} +q^{'}$ constitutes the
``quark sea'' \cite{cheng,johan,song,hd,hdcharm} and the effective  Lagrangian for the {\it
	SU}(4) $\chi$CQM describing interaction between
quarks and 16 GBs can be
expressed as ${{\cal L}} = g_{15}{\bar \psi}\left( \Phi' \right)
{\psi} \,. $ Here $g_{15}$ is the coupling
constant, $\psi$ is the quark field 
and $\Phi'$is the GB field expressed as 
\bea \footnotesize \left( \ba{ccccc} \frac{\pi^0}{\sqrt 2}
+\beta\frac{\eta}{\sqrt 6}+\zeta\frac{\eta^{'}}{4\sqrt
	3}-\gamma\frac{\eta_c}{4} & \pi{^+} & \alpha K{^+} & \gamma
\bar{D}^0\\ \pi^- & -\frac{\pi^0}{\sqrt 2} +\beta
\frac{\eta}{\sqrt 6} +\zeta\frac{\eta^{'}}{4\sqrt 3}
-\gamma\frac{\eta_c}{4}& \alpha K^0 & \gamma D^-\\ \alpha K^- &
\alpha \bar{K}^0 &- \beta \frac{2\eta}{\sqrt 6} +
\zeta\frac{\eta^{'}}{4\sqrt 3}- \gamma\frac{\eta_c}{4} & \gamma
D^-_s\\ \gamma D^0  &\gamma D{^+}& \gamma D^+_s&-
\zeta\frac{3\eta^{'}}{4\sqrt 3}+ \gamma\frac{3\eta_c}{4} \ea
\right)\,. \nonumber\eea 

The spin structure of the spin $\frac{1}{2}^+$  charmed baryon multiplets are defined as
\cite{cheng,johan,hd} $ \widehat B^{\frac{1}{2}^+} \equiv \langle B^{\frac{1}{2}^+}(p', J_z')|{\cal N}|B^{\frac{1}{2}^+}(p, J_z)
\rangle\,,  $ where $|B^{\frac{1}{2}^+}\rangle$ is the  wave
function for the spin $\frac{1}{2}^+$  baryon multiplet and  ${\cal N}$ is the  number operator defined in terms of the number of $q_{\pm}$ quarks.
For each constituent quark the substitution  made is $ q_{\pm}\rightarrow \sum P_{[q, ~ GB]} q_{\pm}+ |\psi(q_{\pm})|^2$,
where $\sum P_{[q, ~ GB]}$ is the probability of emission of
GBs from a $q$ quark and $|\psi(q_{\pm})|^2$ is the transition probability of
fluctuation of every constituent $q_{\pm}$ quark  \cite{hdcharm}.
In general, the wave function for the three-quark system made from any of
the $u$, $d$, $s$, or $c$ quarks is given as a product of $\phi \chi \psi$, with $\phi$ being the
flavor part, $\chi$ the spin part and $\psi$  the
spatial part of the wave function \cite{nscharm,yaoubook}.
In the present work, we will use the set of parameters which have already been fixed in the context of the calculations of  spin and flavor distribution functions\cite{hdcharm,nscharm}. We have $ \phi =20^o\,,~~a=0.12\,,~~ a \alpha^2 \simeq a \beta^2=
0.0243\,,~~ a \zeta^2 = 0.0053 \,, ~~ {\rm and} ~~
a \gamma^2=0.0014\,.$

\section{$Q^2$ evolution and axial-vector charges for quarks}

\begin{figure}
	\includegraphics[width=4in] {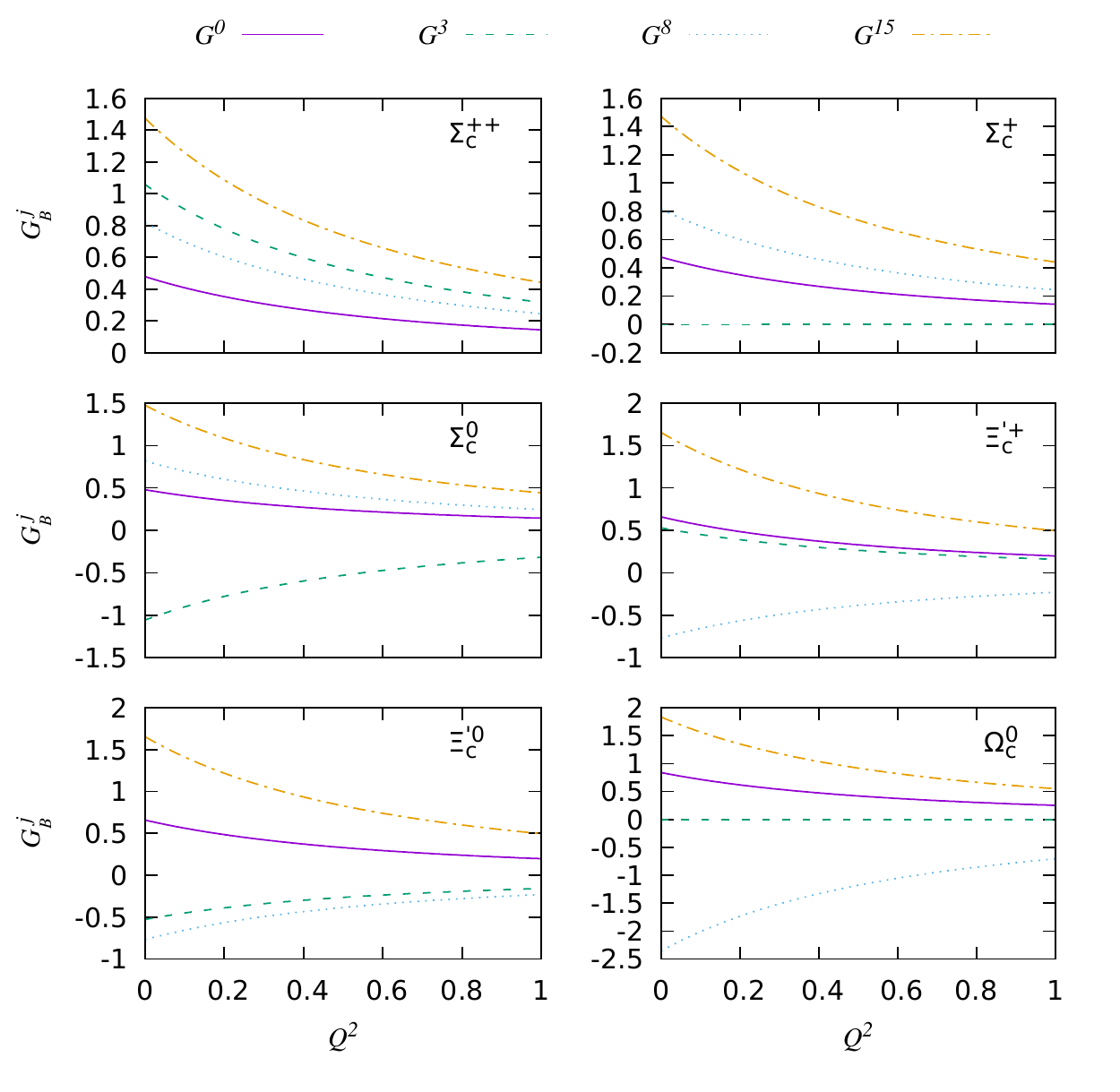}
	\caption{The axial-vector charges for  the spin $\frac{1}{2}^+$ mixed symmetry 20-plet with  the singly charmed sextet (6,1).}
	\label{Spin_1by2_Sextet}
\end{figure}

\begin{figure}
	\includegraphics[width=4in] {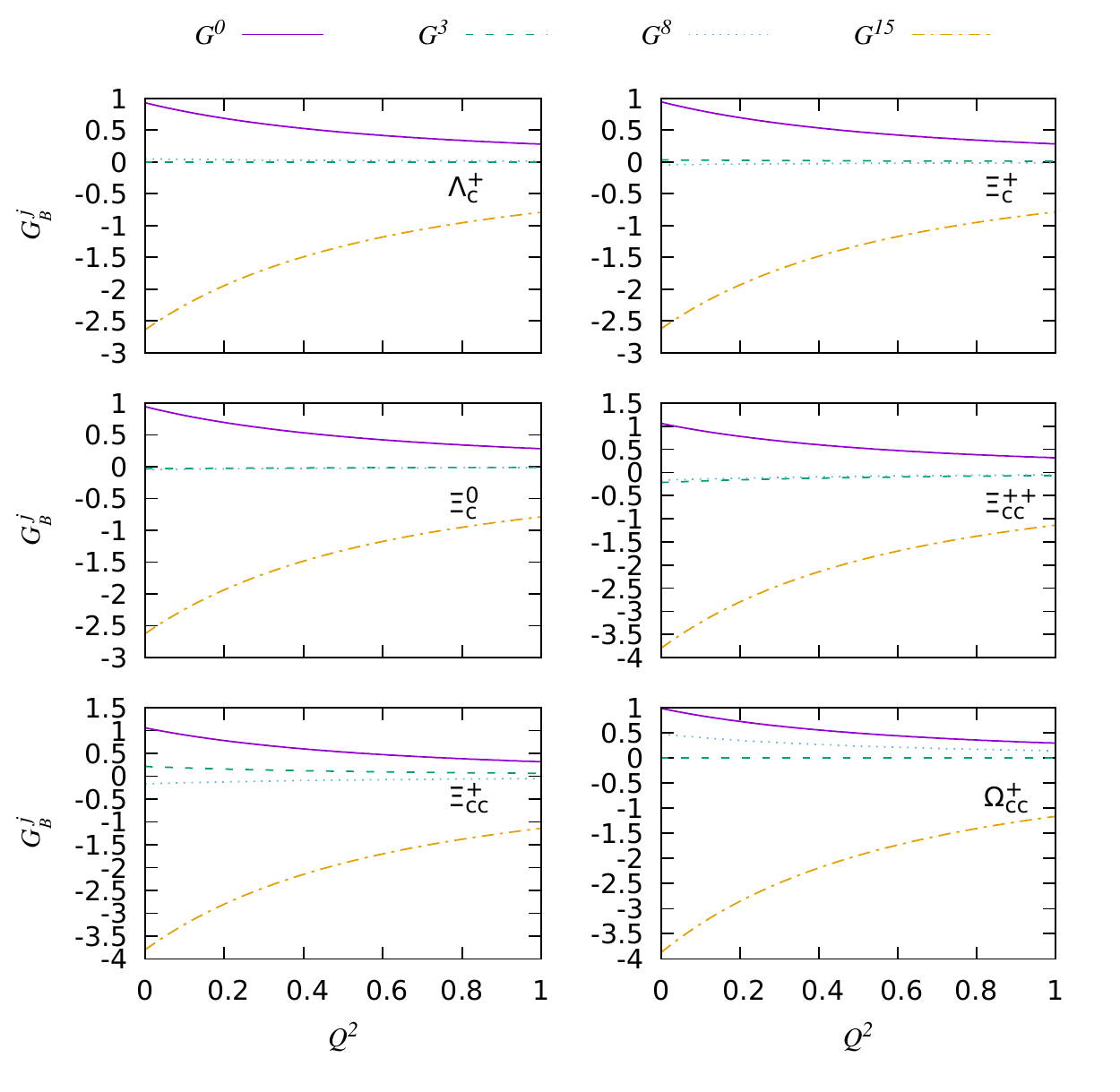}
	\caption{ The axial-vector charges for  the spin $\frac{1}{2}^+$ mixed symmetry 20-plet with the anti-triplet ($\bar{3}$,2) and triplet (3,2) with doubly charmed baryons.}
	\label{Spin_1by2_Triplet}
\end{figure}

The $Q^2$ evolution of the axial-vector charges of the light baryons has been experimentally investigated from the quasi elastic neutrino scattering \cite{antineutrino1,antineutrino2} and from the pion electroproduction \cite{pion-electro} in the past few years.  The most widely used dipole form of parametrization used to analyze the axial-vector charges where the axial mass are extracted from the neutrino-nuclei scattering experiment. In Fig. \ref{Spin_1by2_Sextet}, we have presented the axial-vector charges for  the spin $\frac{1}{2}^+$ mixed symmetry 20-plet with the singly charmed sextet (6,1)  plotted as function of  $Q^2$. 
The singlet and charmed charges in this case have different values because of a charm quark in the constituent structure leading to a significant charm spin polarization. Depending on the magnitude at $Q^2=0$ all the axial-vector charges decrease with increase in the value of $Q^2$. Higher the magnitude at $Q^2=0$, more rapid is the change. 
For the spin $\frac{1}{2}^+$ mixed symmetry 20-plet anti-triplet ($\bar{3}$,2) and triplet (3,2) with doubly charmed baryons presented in Fig. \ref{Spin_1by2_Triplet}, the value of the charmed charge dominates over singlet, triplet and octet charges. The constituent structure includes  two charm quarks and the charm spin polarization is expected to be large.   Lattice calculations and future experiments  at EIC will not only have the possibility to illuminate the complicated issue of spin structure but also impose significant and decisive restraints in different kinematic regions. 

%
%

\end{document}